\documentclass[twocolumn,showpacs,prb,superscriptaddress,aps,floatfix]{revtex4-1}

\usepackage{rotating}
\usepackage{amsmath}
\usepackage{amssymb}
\usepackage{color}
\usepackage{graphicx}
\usepackage[active]{srcltx}
\usepackage[sort&compress]{natbib}
\usepackage{amssymb}
\usepackage[dvips]{epsfig}
\usepackage{float}
\usepackage{braket}
\usepackage{slashed}
\usepackage{tikz}
\usetikzlibrary{positioning,shapes,shadows,arrows}
\usepackage{verbatim}
\usepackage{mathtools}
\usepackage{dsfont}

\tikzstyle{element} = [rectangle, rounded corners, minimum width=3cm, minimum height=1cm,text centered, draw=black, fill=yellow!20]
\tikzstyle{line} = [draw, -latex']

\newcommand{\be}{\begin{equation}}
\newcommand{\ee}{\end{equation}}
\newcommand{\bea}{\begin{eqnarray}}
\newcommand{\eea}{\end{eqnarray}}

\def\pp		{{\bf p}}
\def\rr		{{\bf r}}

\def\AA		{{\bf A}}

\def\QQ		{{\bf Q}}
\def\GG		{{\bf G}}

\def\qq		{{\bf q}}
\def\kk		{{\bf k}}

\def\jj		{{\bf j}}

\def\ee         {\mathbf e}

\def\ga         {\alpha}

\def\gc         {\gamma}

\def\gd         {\delta}
\def\gD         {\Delta}
\def\gee        {\epsilon}

\def\go         {\omega}
\def\gO         {\Omega}

\renewcommand{\[}{\left[}
\renewcommand{\]}{\right]}
\renewcommand{\(}{\left(}
\renewcommand{\)}{\right)}

\def\nk         {n{\bf k}}

\def\bverb      {\renewcommand{\baselinestretch}{1}\begin{verbatim}}
\def\everb  	{\end{verbatim}\renewcommand{\baselinestretch}{1.3}}

\def\ai         {{\it ab-initio}}
\def\ga         {\alpha}

\def\gD         {\Delta} 
\def\gee        {\varepsilon}

\def\go         {\omega}

\def\kk         {{\mathbf k}}
\def\qq         {{\mathbf q}}
\def\rr         {{\mathbf r}}

\def\nk         {n\kk}

\renewcommand{\[}{\left[}
\renewcommand{\]}{\right]}
\renewcommand{\(}{\left(}
\renewcommand{\)}{\right)}

\newcommand{\cnr} {Istituto di Struttura della Materia of the National Research Council, Via Salaria Km 29.3,I-00016 Monterotondo Stazione, Italy}
\newcommand{\etsf} {European Theoretical Spectroscopy Facilities (ETSF)}
\newcommand{\cfc} {CFisUC, Department of Physics, University of Coimbra, Rua Larga, 3004-516 Coimbra, Portugal}

\begin{document}

\title{Lamb shift of the Dirac cone of graphene}

\author{Pedro Miguel M. C. de Melo}
\affiliation{\cnr}
\affiliation{\cfc} 

\author{Andrea Marini}
\affiliation{\cnr} 
\affiliation{\etsf} 

\date{\today}
\begin{abstract}
The fluctuations of the electromagnetic vacuum are one of the most powerful manifestations of the quantum structure of nature.
Their effect on the Dirac electrons of graphene is known to induce some spectacular and purely
quantistic phenomena, like the Casimir and the Aharanov--Bohm effects. 
In this work we demonstrate, by using a first principles approach, that the Dirac cone of graphene is also affected by a sizable Lamb
shift.
We show that the microscopic electronic currents flowing on the graphene plane
couple efficiently with the vacuum fluctuations causing a renormalization  of the electronic levels (as large as 4 meV) and of the velocities.
This Lamb shift is one order of magnitude larger than the value predicted for an
isolated carbon atom.
\end{abstract}      






\maketitle

\label{intro}
Graphene, a two dimensional hexagonal lattice composed of carbon atoms, has become one of the most intensively studied materials in recent years due to its
potential applications in technology~\cite{Geim2007} and in many different fields of theoretical and computational Physics and
Chemistry~\cite{PhysRevLett.108.173601}. Graphene based research is opening new possibilities in the field of optoelectronics due to its unique electronic
properties~\cite{PhysRevLett.108.173601} (e.g. high carrier mobility, ability to tune the charge density) and it is becoming a contender to replace silicon
based electronics, as the latter approaches its nanoscale limits~\cite{Geim2007}.

An intriguing aspect of graphene is represented by the interaction of its Dirac cone electrons with the quantized electromagnetic field. This is a peculiar
effect as most of the equilibrium~\cite{Onida2002}, as well as out--of--equilibrium~\cite{Stefanucci2013,Haug2008} physics studied  in semiconductors and
nano--structures rely on a classical description of the external electric and magnetic fields. This assumption is motivated by the use of low--intensity fields,
well described within a classical framework.

Nevertheless, the interaction of electrons with quantized magnetic fields is the driving mechanism, among others, of the of Aharonov--Bohm\,(AB)\cite{Aharonov1959} effect
and the Casimir force~\cite{Fialkovsky2009,Fialkovsky2011}.

The AB effects is a purely quantum mechanical phenomena which does not have a counterpart in classical mechanics. 
Quantum mechanics, indeed, predicts that a magnetic field $\mathbf{B}$ confined in a closed region inside a carbon nanotube will alter the kinematics of the electons 
traveling around the tube~\cite{Charlier2007,Sangalli2011}. This is the AB effect that
disappears when the external magnetic field is removed. Nevertheless, even in the case where no external magnetic and electric field are present,
the electromagnetic field vacuum is still characterized by a finite, non vanishing energy: the zero--point energy (ZPE). This energy is equivalent to the ZPE
of a quantistic harmonic oscillator. In the case of the electromagnetic field the oscillators are the photons. 

A known manifestation of the ZPE of the electromagnetic field is the Casimir force.
If we take two neutrally charged bodies and place them at close distance, the electromagnetic ZPE will lead to the appearance of an
additional force between them, the Casimir force. 
The strength of the Casimir effect on graphene has been studied using quantum field theory together
with a Dirac model used to describe the $\pi$/$\pi^*$ bands~\cite{Fialkovsky2009,Fialkovsky2011}. 
Surprisingly, a graphene monolayer placed parallel to a
perfect flat conductor exhibits stronger Casimir forces at large separation distances and temperature than what was expected for a material as thick as an
atom~\cite{Fialkovsky2011}\footnote{The intensity of the Casimir force between two graphene sheets rapidly decreases when we reach the micrometer scale and
approaches an intensity of a thousandth of the predicted value for two infinite parallel perfect conducting planes placed at the same
distance~\cite{Drosdoff2010,Bordag2001}}.

The Lamb shift affects even a single electron immersed in the vacuum of a quantized electromagnetic field and it does not require the presence of substrate or
of an external perturbation. 
The original observation of the Lamb shift in the Hydrogen atom has represented a corner--stone in the development of quantum electrodynamics. Nevertheless
it is still an active field of research~\cite{Lindley2012,Scully2010} especially in the case of N--atoms systems where the Lamb shift is expected to be
amplified by cooperative effects. It has been observed~\cite{Dicke1954}, indeed, that the presence of additional proximate atoms changes the strength of the
interaction with the virtual photons. This cooperative Lamb shift gives rise to superradiance phenomena and corrections to the energy levels of the compound
system\cite{Keaveney2012,Gramich2014}. Superradiance effects have been observed, for example, in atomic vapours~\cite{PhysRevLett.108.173601} and mesoscopic
atomic arrays~\cite{Meir2014}.

A crystalline solid is a simple but clear example of a N--atoms system. Thus we expect that the cooperative emission of virtual photons is amplified. If we add
this simple argument to the special interaction of the Dirac electrons with the quantized electromagnetic field we are led to the conclusion that the Lamb
shift in graphene can be sizable.

In this work, indeed, we use an  \ai\, and atomistic approach to calculate the effect of the interaction of electrons with the quantized electromagnetic vacuum.
The problem is rewritten in a Kohn--Sham\,(KS) basis and the electron--photon interaction Hamiltonian is expanded in plane--waves. We show that this interaction
leads to a Lamb shift of the energy levels of graphene as big as $-4$\,meV and a consequent reduction of its electronic speed. We show that only by using an \ai\,
approach it is possible to describe the multitude of states involved in the virtual transitions caused by the Lamb shift. In addition we will discuss that,
physically, the microscopic mechanism that drives the Lamb shift is the interaction of the electromagnetic field with the microscopic currents caused by the
material spatial discontinuity. These currents represent an intrinsic property of any material and, in general, they can be excited only by using an external
perturbation. The Lamb shift is a striking quantistic manifestation of their intrinsic existence. 


In order to describe the effect induced by the electromagnetic vacuum fluctuations we first introduce the Hamiltonian describing the interaction of the
paramagnetic current $\hat{\jj}\(\rr\)$ with an electromagnetic field, $\hat{\AA}\(\rr\)$:
\begin{align}
\hat H_{e-\gc} = -\frac{1}{c} \int d\rr\, \hat{\AA}\(\rr\)\cdot \hat{\jj}\(\rr\).
\label{eq:1}
\end{align}
We are interested in calculating the effect of $H_{e-\gc}$ in extended systems. Therefore we use a super--cell approach where the  atomic lattice is described
by a volume $\gO$ repeated periodically. As a consequence, in order to introduce the second quantization for the vector potential, we Fourier expand it: 
\begin{align}
\hat{\AA}\(\rr\) = 
\sum_{\ga}\int\,\frac{d\QQ}{\(2 \pi\)^3} \sqrt{\frac{2\pi c^2 \gO}{\go_{\QQ}}}
\[\hat{d}_{\QQ\ga}e^{i \QQ\cdot\rr} +\text{H.c.}\] \ee_{\QQ\ga},
\label{eq:2}
\end{align}
with $\QQ\equiv \qq+\GG$ a generic point in the reciprocal space composed by a sum of the vector inside the Brillouin Zone ($\qq$) and a vector of the reciprocal lattice ($\GG$). 

In Eq.\eqref{eq:2} $\ee_{\QQ\ga}$\footnote{Together with the photon momentum $\QQ$ the vectors $\ee_{\QQ\ga}$ form an orthogonal basis. The form of
$\ee_{\QQ\ga}$ is dependent on the polarization of light: for linear polarization $\ee_{\QQ\ga}=\ee_{x/y}$; for circular polarization $\ee_{\QQ\ga} =
\cos\theta\ee_x \pm \sin\theta\ee_y$. In the derivation of Eq.\eqref{eq:4} we assume the photon momentum to be aligned with the z Cartesian direction.} is the
photon polarization vector relative to the photon branch $\alpha$ and orthogonal to $\QQ$ (we are working in Coulomb's gauge, so $\nabla\cdot\AA\(\rr\) ={\bf
0}$). $\go_{\QQ}=c |\QQ|$ is the energy of the free~\footnote{We neglect renormalization effects of the photons, that are assumed to be the quanta of a free
field. This approximation is well motivated for an isolated graphene sheet as the photons are not spatially confined and the electronic screening can be safely
assumed to be negligible.} photon whose creation and annihilation operators are $\hat{d}_{\QQ\ga}$ and $\hat{d}^{\dagger}_{\QQ\ga}$.

In this work we treat the electrons with an \ai\, approach. This is advantageous as the \ai\, framework represents the most up--to--date and accurate method to
describe the atomic structure of realistic materials. This approach aims at modeling the ground and excited states of realistic materials and it is incessantly
developed thanks to the use of combined numerical and theoretical accurate approaches. This is largely possible thanks to the description of the atomistic
properties of the materials by means of Density--Functional--Theory\,(DFT)\cite{R.M.Dreizler1990}. DFT is an \textit{ab-initio} ground--state theory that allows
to calculate \textit{exactly} the electronic density and the total energy without relying on any adjustable parameter. Thanks to its wide spread use, DFT is now
a standard tool commonly used in the vast material science community. As it will be discussed shortly the ability of DFT to calculate all possible empty states
will be crucial in describing the wealth of virtual states involved in the transitions that cause the Lamb shift. By using a low--energy tight--binding or Dirac
model this would not be possible. 

Thusly, the paramagnetic electronic current density operator, which is defined as $\hat{\jj}\(\rr\)\equiv \frac{-i}{2}\[\hat\psi^\dagger\(\rr\)\nabla_{\rr}\hat\psi\(\rr\) - \text{H.c.}\]$, can be easily rewritten in the KS basis by expanding the field operators, $\hat{\psi}\(\rr\)=\sum_{\nk}\phi_{\nk}\(\rr\)\hat{c}_{\nk}$. Here $\phi_{\nk}\(\rr\)$ is a KS orbital with energy $\gee^{KS}_{\nk}$ and creation and annihilation operators $\hat{c}_{n\kk}$ and $\hat{c}^{\dagger}_{n\kk}$. It follows that:
\begin{align}
\hat{\jj}\(\rr\)=\frac{1}{2}\sum_{nm,\kk\pp} \[\hat{c}^{\dagger}_{\nk} \hat{c}_{m\pp} \pp^{\nk}_{m\pp}\(\rr\)+\text{H.c.}\],
\label{eq:3}
\end{align}
with $p^{\nk}_{m\pp,i}\(\rr\)\equiv \(-i\) \phi^*_{\nk}\(\rr\)\(\partial_{r_i}\phi_{m\pp}\(\rr\)\)$. Thanks to Eq.\eqref{eq:3} it follows that all ingredients of our approach are calculated \ai, in a parameter--free way.

Now that the interaction Hamiltonian is defined we can apply standard perturbation theory. At difference with the electron--electron or electron--phonon cases,
the electron--photon interaction can be treated perturbatively as it strength is dictated by the small hyperfine constant $\ga=\frac{1}{137}$. It is indeed well
known that the first non vanishing order in the perturbation expansion of the energy correction in powers of $H_{e-\gc}$ accounts for the biggest part of the
correction to the energy levels\cite{Shabaev2002}. 

It is, then, straightforward to show that the second order contribution to the renormalization of the level $\nk$ is $\gee_{\nk}=\gee^{KS}_{\nk}+\gD
\epsilon_{\nk}\(\{N_{\QQ\ga}\}\) + i\Gamma_{\nk}$, with $\{N_{\QQ\ga}\}$ the occupation factors of the photon population, while $\gD\epsilon_{\nk}$ is the
single particle energy shift and $\Gamma_{\nk}$ represents its lifetime. The Lamb shift  appears when we assume that there are no photons and $N_{\QQ\ga}=0$.
In this limit $\Delta\gee_{\nk}$ is not zero and provides the zero--point correction to the energy levels:
\begin{multline}
\epsilon_{n\kk} -\epsilon_{\nk}^{KS}= 
\sum\nolimits'_{m}\int \frac{d\QQ}{(2\pi)^3}
\frac{\pi\[\sum_{ij}\tau^{\QQ}_{ij} P_{nm\kk,i}^{\QQ} \(P_{nm\kk,j}^{\QQ}\)^*\]}
{2\go_{\QQ} }\\
\times\[\frac{1 - f_{m}(\kk-\qq)}{\gee^{KS}_{\nk}-\gee^{KS}_{m\kk-\qq}-\go_{\QQ} + \mathrm i 0^+} + \right.\\
\left.\frac{f_{m}(\kk-\qq)}{\gee^{KS}_{\nk}-\gee^{KS}_{m\kk-\qq}+\go_{\QQ} - \mathrm i 0^+}\].
\label{eq:4}
\end{multline}
In Eq.\eqref{eq:4}, $\sum_{m}\nolimits'$ excludes the contribution $m=n$ when $\QQ\rightarrow{\bf 0}$. We have also defined 
\begin{align}
P_{nm\kk,i}^{\QQ} = \int d\rr\,e^{i \rr\cdot\QQ}\[\pp^{n\kk}_{m\kk-\qq,i}\(\rr\)+\text{H.c.}\],
\label{eq:5}
\end{align}
and the transverse matrix $\tau^{\QQ}_{ij}= \sum_{\ga}e_{\QQ\ga i}e_{\QQ\ga j}$\footnote{It is straightforward to show that $\tau^{\QQ}_{ij}=
\sum_{\ga}e_{\QQ\ga i}e_{\QQ\ga j} = \delta_{ij} - \frac{Q_iQ_j}{Q^2}$, where $\delta_{ij}$ is the Kronecker delta function. This matrix eliminates all
components which are parallel to the photon momentum $\QQ$.}. Using Eq.~\eqref{eq:4} we can also easily evaluate the change in the electronic velocity $\gd
v_{\nk}= \frac{v_{\nk} - v_{\nk}^\text{KS}}{v_{\nk}^\text{KS}} = \frac{\partial \gD\epsilon_{\nk}}{\partial \epsilon_{\nk}^\text{KS}}$, as explained in Supplemental Material.

From Eq.~\eqref{eq:4} we notice that, as $\go_\QQ$ is linear in $|\QQ|$, $\gD_{\nk}$ is divergent for $\QQ\to\infty$. As explained in the Supplemental Material. the
regularization of $\gD_{\nk}$ is obtained by splitting the real part of Eq~\eqref{eq:4} in the $\GG=0$ and $\GG\neq0$ parts, so that we can define
\begin{align}
\gD\epsilon_{\nk}|_\text{regular} = \gD\epsilon_{\nk}|_{\GG=0} + 2\gD\epsilon_{\nk}^\text{filled}|_{\GG\neq0},
\end{align}
with $\gD\epsilon_{\nk}^\text{filled}|_{\GG\neq0}$ corresponding to the real part of Eq.~\eqref{eq:4}, with the index $m$ running only on filled levels. It is straightforward to prove that also the electronic speed renormalization, $\delta v_{\nk}$, can be rewritten in a regular way
\begin{align}
\gd v_{\nk}|_\text{regular} = \gd v_{\nk}|_{\GG=0} + 2\gd v_{\nk}^\text{filled}|_{\GG\neq0}.
\end{align}

The graphene ground state has been calculated using the QE~\cite{pwscf} code within the Local Density Approximation\,(LDA)~\cite{R.M.Dreizler1990}.
Eq.\eqref{eq:4} has been implemented in the Yambo code~\cite{Marini20091392}, that is interfaced with QE and can use the calculated KS states. Graphene is
simulated by using a  $24n$ by $24n$ $\kk$ points grid, where $n$ is an integer, with a unit cell with a lattice parameter of 4.60 [a.u.], a layer separation of 14.0 [a.u.]
and a kinetic cut-off of 80 Ry. The resulting corrections to the eigenvalues and electronic speed are then
interpolated with a quadratic polynomial function, using $(24n)^{-1}$ as the independent variable.

\begin{figure}[htpb]
\begin{center}
\includegraphics[width=8cm]{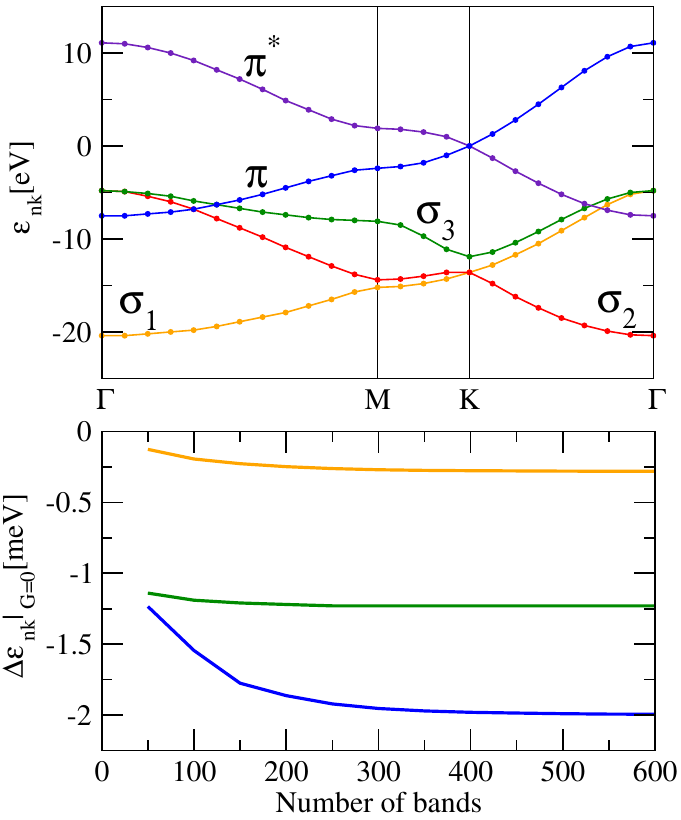}
\caption{\footnotesize{Top: graphene band dispersion for the $\sigma_1$, $\sigma_2$, $\sigma_3$, $\pi$, and $\pi^*$ bands. Bottom: convergence of the energy level correction with $\GG=0$ for the K points of $\sigma_1$, $\sigma_3$, and $\pi$ bands on a 15$\times$15$\times$1\,$\kk$-point grid.}}
\label{fig:4}
\end{center}
\end{figure}

\begin{figure}[htpb]
\begin{center}
\includegraphics[width=8cm]{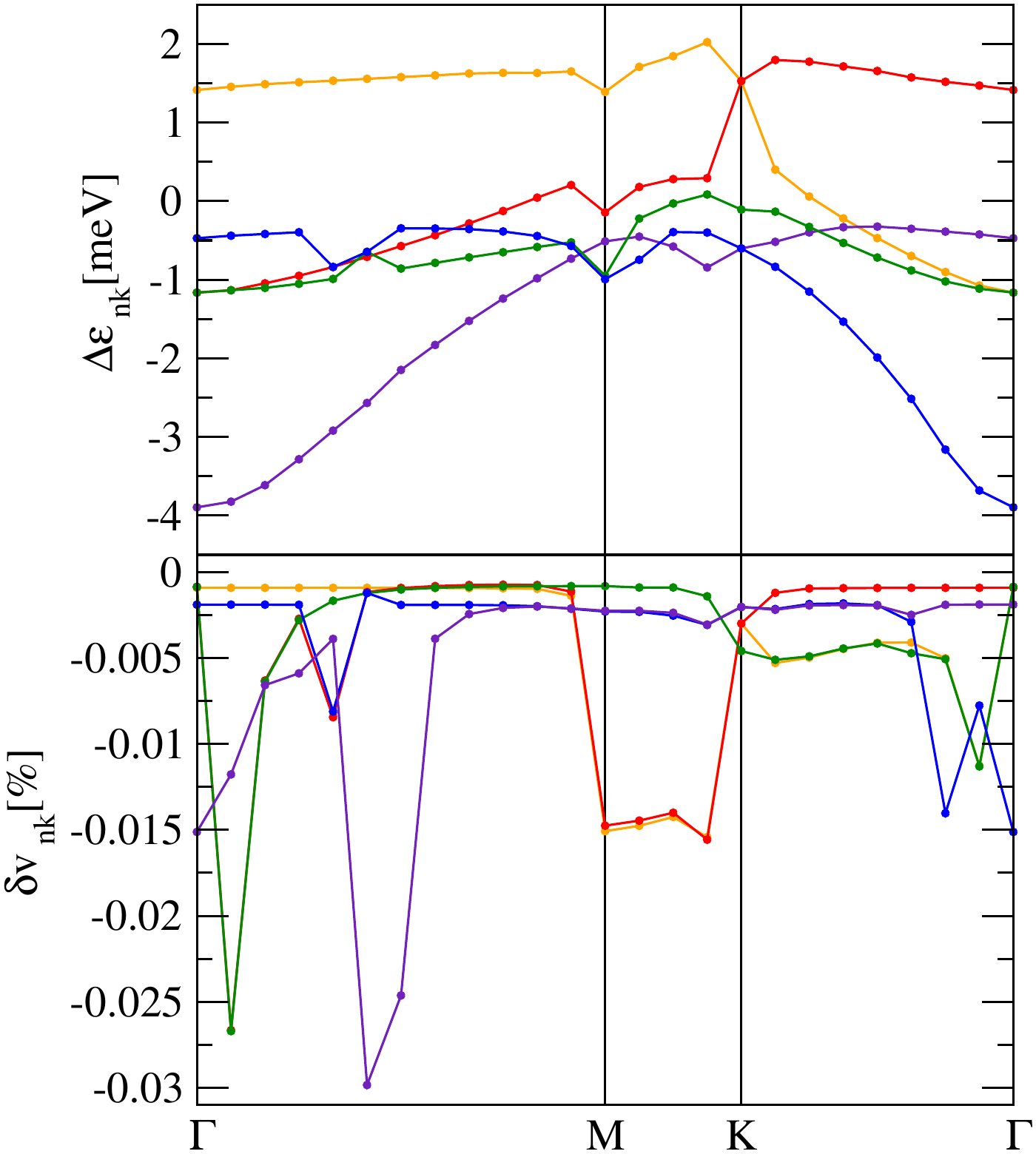}
\caption{\footnotesize{Top: correction of the electronic levels induced by the interaction with the electromagnetic field. Bottom: correction of the electronic speeds induced by the interaction with the electromagnetic field. This different behavior can be explained by looking at the different contributions of these bands to the electronic current.}}
\label{fig:5}
\end{center}
\end{figure}
In Fig.\ref{fig:5} we show the Lamb correction to the energy level (top frame) and velocities (bottom frame). Two aspects are clearly visible. The
first is that the Dirac cone is down--shifted of $0.6$\,meV. This shift is large if compared both to previous calculations and to the case of heavy atoms. Indeed
it has been previously predicted~\cite{Kibis2011} that graphene acquires a band--gap as large as $50$\,$\mu$eV induced by the interaction with the
electromagnetic field fluctuations. In Ref.\cite{Kibis2011} however, graphene is described with a single--band model and, as it will be clear below, this represents an approximation that  dramatically fails in describing the wealth of virtual states involved in the proper evaluation of the Lamb shift. The result we find is two orders of magnitude larger than in this over simplified model.

In addition our correction is large even if compared to the case of heavy atoms. Indeed, for isolated atoms the Lamb shift is known to scale as an $Z^2$. This means that it is, indeed, larger for heavy atoms\footnote{In the case of the muonic deuterium~\cite{Krutov2011} the Lamb shift, $\Delta E$, is as large as $-240\,meV$. Similarly large it is for heavy relativistic atoms~\cite{Shabaev2013} like Roentgenium (Z=272,  $\Delta E\sim-96$\,meV), Caesium (Z=55,  $\Delta E\sim-98$\,meV), E119 ($\Delta E\sim-232$\,meV) and E120 ($\Delta E\sim-250$\,meV).}. These corrections approximatively follow a simple rule: $\Delta E\sim \Delta E_0 Z^2$ with $\Delta E_0\approx\,16.86\mu$eV that would imply, for a Carbon atom, a $\Delta E_0\approx\,0.6$\,meV, one order of magnitude smaller that
the value we found.

As graphene is composed only by light Carbon atoms the reason for such a large Lamb correction must be searched elsewhere and not in arguments based solely on the atomic number. Equation \eqref{eq:4} describes $\Delta_{\nk}$ as a process where the initial $\ket{\nk}$ undergoes virtual transitions to all possible states $\ket{m\kk-\qq}$ emitting a photon $\ket{\QQ,\ga}$.  The creation of a virtual population of photons that is annihilated at the end of the process, is an alternative physical picture of the Lamb shift that makes clear its link with the potential space of final states that can be reached.

The dimension of this space represents the multitude of potential final states. This multitude is reminiscent of the multitude of charges that characterize the collective Lamb shift~\cite{Dicke1954} and suggests that, indeed, the shift can be large in extended systems.

In order to highlight the crucial importance of the space spanned by the virtual transition in Eq.\eqref{eq:4} we show, in the bottom panel of
Fig.\ref{fig:4}, the contributions of the different bands to the Lamb shift of the Dirac point. 

Another important aspect that emerges clearly from our calculations is that the $\sigma$ bands are also affected by a correction that is smaller (even if of the same order of magnitude) of the one relative to the Dirac cone. The difference between the $\sigma$ bands down-shift and the shift of the $K$ point results in a reduction of the occupied band with of $2.13$\,meV. 

From Fig.\ref{fig:5} we see that the renormalization of the velocities is one order of magnitude smaller ($\sim 5\%$ versus $\sim 0.01\%$) compared to the case of the electron--phonon interaction~\cite{Giustino2007,Park2008}. Although small (bellow 1\%) the electronic speed renormalization follows the very same trend of the electronic levels, with the $\pi$ and $\pi^*$ bands more affected than the $\sigma$ bands. On the other hand the large reduction of the band--width $W$ points to a reduction of the hopping assisted transport with potential implications on the transport properties. 

\begin{figure}[htpb]
\begin{center}
\includegraphics[width=8.3cm]{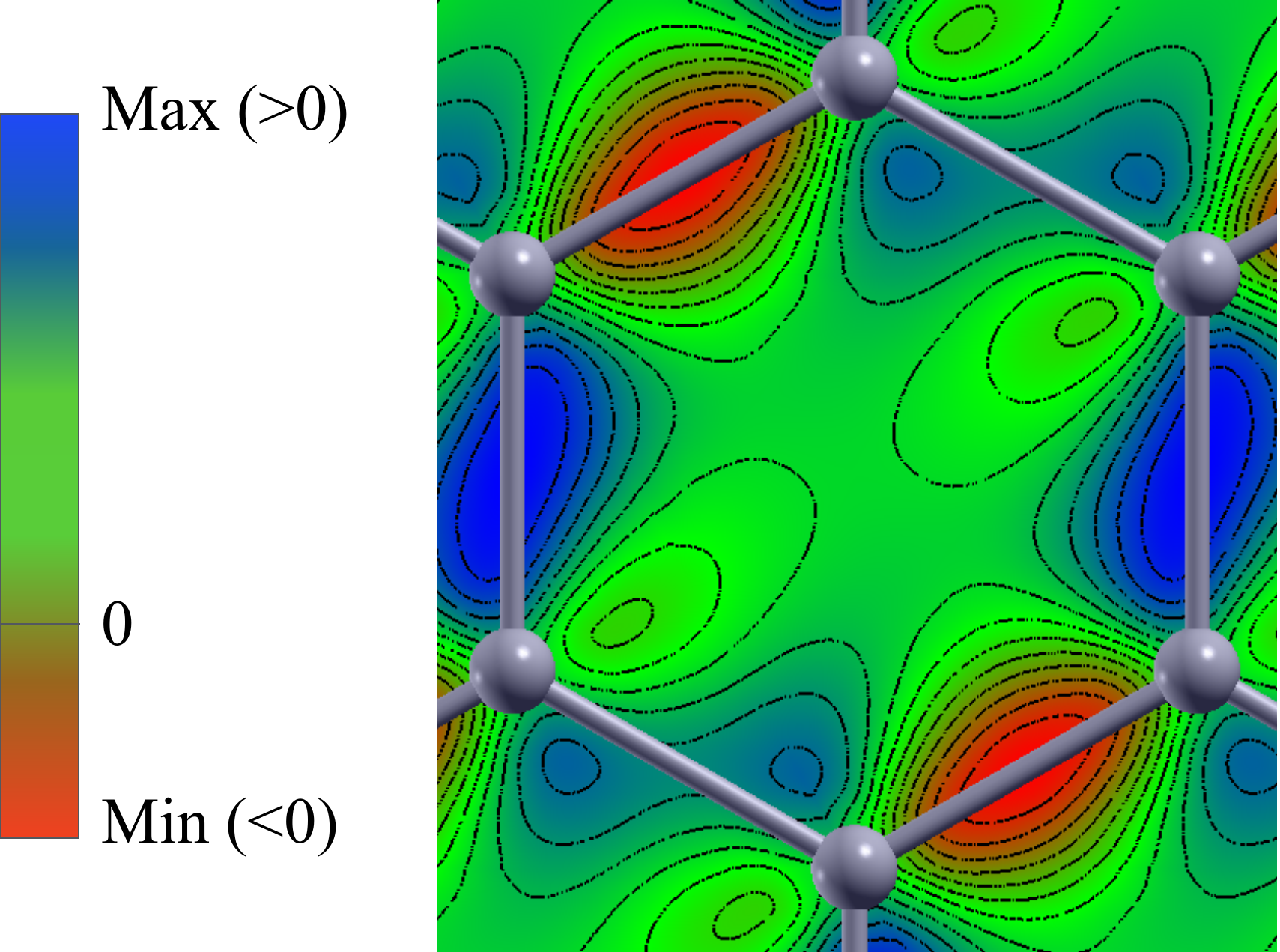}
\caption{\footnotesize{Two--dimensional plot of the current $\jj\(\rr\)$. This is dictated by the spatial anisotropy of the KS orbitals and, indeed, it has the same symmetry of the lattice. }}
\label{fig:4a}
\end{center}
\end{figure}

In order to pin down the physical motivation of this large Lamb shift we propose an alternative approach. Indeed, the Lamb shift is historically connected with the virtual emission and absorption of photons. This picture is a direct consequence of the perturbative treatment of the interaction Hamiltonian, Eq.\eqref{eq:1}. Nevertheless, an alternative microscopic interpretation, is based on the classical picture of the interaction of electrical currents with the
electromagnetic field. This interaction is at the basis of the Lorentz force, for example, and it is linked with the form of the interaction Hamiltonian.

Even if there are no external fields any material is crossed by microscopic currents caused by the intrinsic spatial discontinuity of the material. Indeed, even
if for continuous materials $\braket{\jj\(\rr\)}={\bf 0}$, for systems like graphene a finite current flows between regions with different density. On the
average these currents do not produce a macroscopic current but, nevertheless, do interact with the electromagnetic field. This microscopic interaction is the
source of the Lamb shift.

The projection of the electronic current $\jj(\rr)$ along the graphene plane is shown in Fig.\ref{fig:4a}.

\begin{figure}[htpb]
\begin{center}
\includegraphics[width=8.3cm]{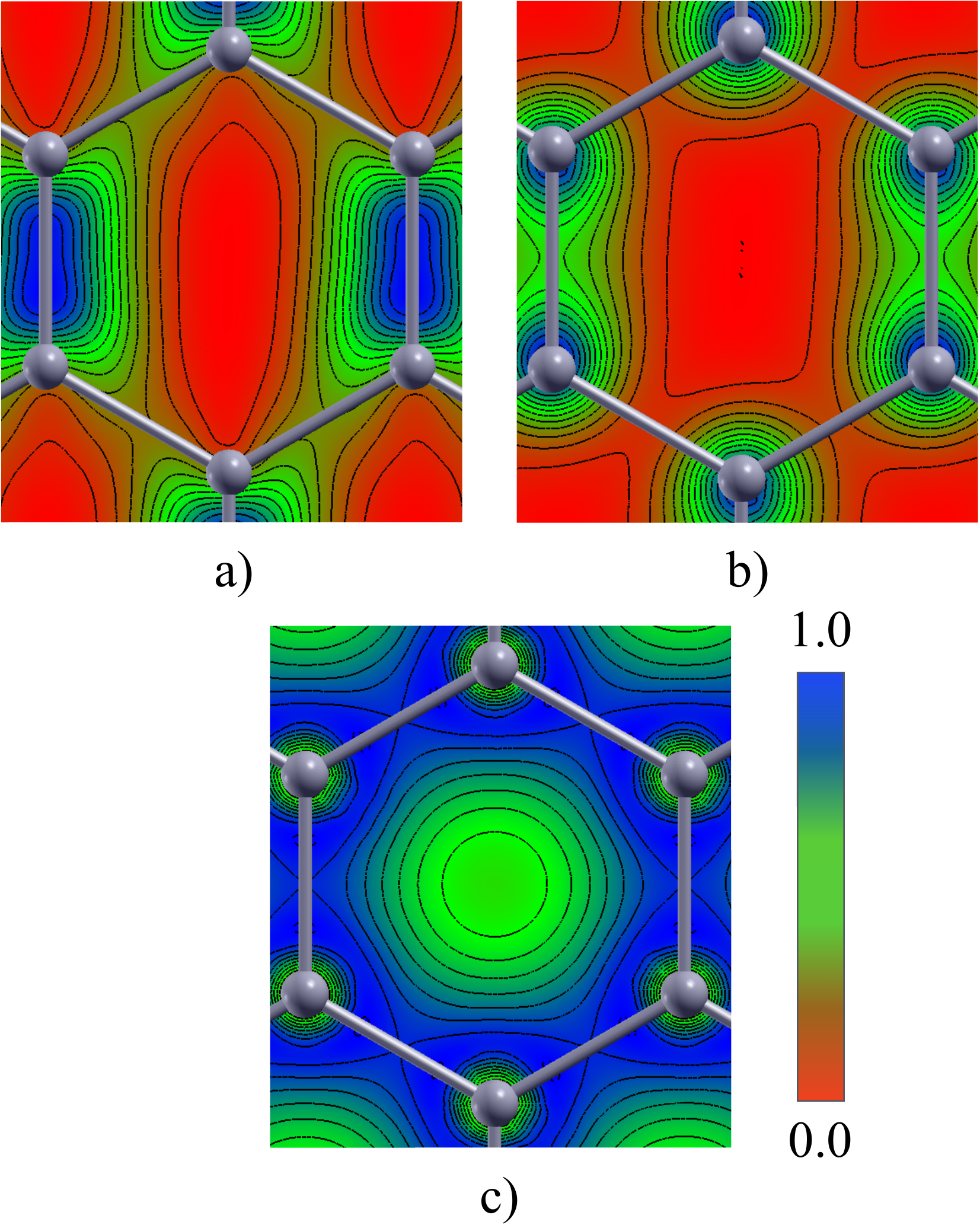}
\caption{\footnotesize{ a), b) and c) are, respectively, the wave function amplitudes of bands $\sigma_2$, $\pi^*$, and $\sigma_3$ at the K point on the graphene plane. We can see that the microscopic currents flow between regions of different density and mainly along the hexagonal edges. These fluctuations are enhanced by the  $\sigma_2$ and$\pi^*$ states while are washed out by the more uniform $\sigma_3$ state. Hence the higher energy correction shown in fig.\ref{fig:5}. On the average, of course, this microscopic quantity does not produce a macroscopic current. Nevertheless these microscopic currents couple with the microscopic vector--potential in the same way of a macroscopic current generated by an external electromagnetic field.}} \label{fig:4b}
\end{center}
\end{figure}

Here we can clearly see that the microscopic current is mostly zero, with the non--zero regions concentrated along the sides of the graphene hexagon. Therefore, these
current fluctuations mainly interact with states which wave function amplitudes oscillates along the hexagonal edges of the unit cell alternating positive to
negative regions. As exemplified in Figure~\ref{fig:4b}, the $\pi$ and $\pi^*$ states, as well as some $\sigma$ bands, show this alternating behaviour and,
therefore, more strongly interact with the vacuum fluctuations. The more $s$--like orbitals, instead, have a uniform spatial distribution (see the $\sigma_3$
state shown Figure~\ref{fig:4b}) and their Lamb shift  is smaller.

In conclusion, using an \ai\, approach, we have predicted that the Dirac cone of graphene is characterized by a sizable Lamb shift induced by the interaction of
the massless Dirac electrons with the vacuum fluctuations. We predict this shift to be larger of what is expected from semi--empirical model calculations and
also if compared to the case of heavy isolated atoms. This is explained in terms of cooperative effects caused by the presence of a multitude of atoms. Moreover
we trace back the different corrections for $\sigma$ and $\pi$ bands to their peculiar contribution to the microscopic currents flowing in the material. The
present results do contribute to improve the state-of-the-art understanding of the Lamb shift in realistic and extended materials. They clearly bind a
quantitative description of the shift to to a careful and parameter-free description of the full spectrum of the graphene electronic states.

\subsection*{Acknowledgements}

We acknowledge financial support by the {\em Futuro in Ricerca} grant No. RBFR12SW0J of the Italian Ministry of Education, University and Research MIUR. AM also
acknowledges the funding received from the European Union project MaX {\em Materials design at the eXascale} H2020-EINFRA-2015-1, Grant agreement n. 676598 and
{\em Nanoscience Foundries and Fine Analysis - Europe} H2020-INFRAIA-2014-2015, Grant agreement n. 654360. PM thanks the Portuguese Foundation for Science and
Technology (FCT) and the Portuguese Ministry for Education and Science, for the funding received through the scholarship grant No. SFRH/BD/84032/2012 under the
European Social Fund and the Programa Operacional Capital Humano (POCH).

\bibliographystyle{apsrev4-1}
\bibliography{refs}	

\end{document}


\title{Lamb shift of the Dirac cone of graphene}

\author{Pedro Miguel M. C. de Melo}
\affiliation{\cnr}
\affiliation{\cfc} 

\author{Andrea Marini}
\affiliation{\cnr} 
\affiliation{\etsf} 

\date{\today}

\title{Renormalization of the energy and velocity corrections}

\maketitle

The real part of Eq.\,(4) can be rewritten as
\begin{multline}
\gD\epsilon_{\nk} = \sum\nolimits'_{m}\int \frac{d\QQ}{(2\pi)^3}
\frac{\pi\[\sum_{ij}\tau^{\QQ}_{ij} P_{nm\kk,i}^{\QQ} \(P_{nm\kk,j}^{\QQ}\)^*\]}
{2\go_{\QQ} }\\
\[\frac{1 - f_{m}(\kk-\qq)}{\gee^{KS}_{\nk}-\gee^{KS}_{m\kk-\qq}-\go_{\QQ}} + \frac{f_{m}(\kk-\qq)}{\gee^{KS}_{\nk}-\gee^{KS}_{m\kk-\qq}+\go_{\QQ}}\]\\
= \gD\epsilon_{n\kk}|_{\GG=0} + \gD\epsilon_{n\kk}^\text{filled}|_{\GG\neq 0} + \gD\epsilon_{n\kk}^\text{empty}|_{\GG\neq 0}.
\label{eq:7}
\end{multline}
where $\gD_{\nk}|_{\GG=0}$ is the contribution from all the terms with $\GG=0$
\begin{multline}
\gD\epsilon_{\nk}|_{\GG=0} = \sum\nolimits'_{m}\int_\text{BZ} \frac{d\qq}{(2\pi)^3}\\
\frac{\pi\[\sum_{ij}\tau^{\qq}_{ij} P_{nm\kk,i}^{\qq} \(P_{nm\kk,j}^{\qq}\)^*\]}
{2\go_{\qq} }\\
\[\frac{1 - f_{m}(\kk-\qq)}{\gee^{KS}_{\nk}-\gee^{KS}_{m\kk-\qq}-\go_{\qq}} +\frac{f_{m}(\kk-\qq)}{\gee^{KS}_{\nk}-\gee^{KS}_{m\kk-\qq}+\go_{\qq}}\],
\label{eq:8}
\end{multline}
and $\gD_{n\kk}^\text{filled/empty}|_{\GG\neq 0}$ are, respectively, the contributions of the terms with $\GG\neq 0$ of the filled/empty states. By noticing that in atomic units $c\approx 137$ and, therefore, assuming, in Eq.~\eqref{eq:7}, that $c|\GG| \gg \epsilon_{\nk}^\text{KS} - \epsilon_{m\kk-\qq}^\text{KS}$, we get that
\begin{align}
\gD\epsilon_{\nk}^\text{filled}|_{\GG\neq 0} \approx \sum_{\GG\neq0}\frac{\pi}{2c^2|\GG^2|}R_{\nk,\GG}^\text{filled},
\label{eq:9}
\end{align}
\begin{align}
\gD\epsilon_{\nk}^\text{empty}|_{\GG\neq 0} \approx -\sum_{\GG\neq0}\frac{\pi}{2c^2|\GG^2|}R_{\nk,\GG}^\text{empty}.
\label{eq:10}
\end{align}
The coefficients $R_{\nk \GG}^\text{filled}$ and $R_{\nk \GG}^\text{empty}$ are obtained from the expression
$R_{\nk \GG}^\text{filled} + R_{\nk \GG}^\text{empty} = R_{\nk \GG}$, where
\begin{multline}
R_{\nk,\GG} = 4\sum\nolimits_m'\int\frac{d\qq}{(2\pi^3)}\sum_{\ga\gb}\tau_{\ga\gb}^\QQ\int d\rr d\rr' e^{i(\qq+\GG)\cdot(\rr-\rr')}\\
\braket{\nk|\[r_\ga,H\]|m\kk-\qq}\braket{m\kk-\qq|\[r_\gb,H\]|\nk},
\label{eq:11}
\end{multline}
by restricting the sum over the bands to the filled bands, for the case of $R_{\nk \GG}^\text{filled}$, or to the empty bands, for the case of $R_{\nk \GG}^\text{empty}$. This is a useful work around, as it allows us to replace the term coming from the empty states and write
\begin{align}
\gD\epsilon_{\nk} = \gD\epsilon_{\nk}|_{\GG=0} + 2\gD\epsilon_{\nk}^\text{filled}|_{\GG\neq 0} - \tilde\gD\epsilon_{\nk}|_{\GG\neq 0},
\label{eq:12}
\end{align}
with $\tilde \gD\epsilon_{\nk}|_{\GG\neq 0}$ given by an equation equal to Eq.~\eqref{eq:9} but with $R_{\nk\GG}$ instead of $R_{\nk\GG}^\text{filled}$.
By taking the dipole approximation $\exp\[i\qq\cdot(\rr-\rr')\]\approx 1$ and approximating $\tau_{\ga\gb}^\QQ \approx \tau_{\ga\gb}^\GG$, we can eliminate the sum over the intermediate states, through
\begin{multline}
\sum\nolimits_m'\int\frac{d\qq}{(2\pi)^3}\ket{m\kk-\qq}\bra{m\kk-\qq} = \gd(\rr-\rr') - \\
\int\frac{d\qq}{(2\pi)^3}\ket{n\kk-\qq}\bra{n\kk-\qq},
\label{eq:13}
\end{multline}
and we ignore the contribution from the last term on the right hand side. This allows us to treat the sum over $\GG$ separately
\begin{multline}
4\sum_{\ga\gb}\tau_{\ga\gb}^\GG\braket{\nk|\hat p_\ga\hat p_\gb|\nk} = 4\sum_{\ga\gb} \(\gd_{\ga\gb} - \frac{G_\ga G_\gb}{|\GG|^2}\)\\
\braket{\nk|\hat p_\ga\hat p_\gb|\nk} \\
= 4\sum_\gl |\braket{\gl|\hat \pp|\nk}|^2\(1 - \cos^2\theta_{\nk \gl}^\GG\)
\label{eq:14}
\end{multline}
where $\theta_{\nk \gl}^\GG$ is the angle between $\GG$ and $\braket{\lambda|\pp|\nk}$. Due to the sum over an infinite number of states $\ket{\lambda}$ we can average the value of the cosine 
\begin{multline}
4\sum_\gl |\braket{\gl|\hat \pp|\nk}|^2\(1 - \cos^2\theta_{\nk \gl}^\GG\) \approx 2\sum_\gl |\braket{\gl|\hat \pp|\nk}|^2 = \\
2\braket{\nk|\hat \pp^2 |\nk}.
\label{eq:15}
\end{multline}
This has the advantage of allowing us to approximate the last term in Eq.~\eqref{eq:12} as
\begin{align}
\tilde\gD_{\nk} \approx - \frac{\pi}{c^2}\braket{\nk|\hat \pp^2 |\nk}\sum_{\GG\neq 0}\frac{1}{|\GG|^2},
\label{eq:16}
\end{align}
which is naturally divergent and usually removed from the sum since it corresponds to the self-interaction energy of the bound electron~\cite{Mandl1993}. As such only the first and second terms of Eq.~\eqref{eq:12} are kept. By applying the sample procedure to $\delta v_{\nk}$ we get that
\begin{align}
\delta v_{\nk} = \delta v_{\nk}|_{\GG=0} + 2\delta v_{\nk}|_{\GG\neq0}^\text{filled},
\label{eq:17}
\end{align}
where $\delta v_{\nk}|_{\GG=0}$ is the term with all the $\GG=0$ components, given by
\begin{multline}
\delta v_{\nk}|_{\GG=0}=-\sum\nolimits'_{m}\int \frac{d\qq}{(2\pi)^3}\\
\frac{\pi\[\sum_{ij}\tau^{\qq}_{ij} P_{nm\kk,i}^{\qq} \(P_{nm\kk,j}^{\qq}\)^*\]}
{2\go_{\qq}}\[\frac{1-f_m(\kk-\qq)}{\(\gee^{KS}_{\nk}-\gee^{KS}_{m\kk-\qq}-\go_{\qq}\)^2} +\right. \\
\left. \frac{f_m(\kk-\qq)}{\(\gee^{KS}_{\nk}-\gee^{KS}_{m\kk-\qq}+\go_{\qq}\)^2}\],
\label{eq:18}
\end{multline}
and the contribution from the filled states for which $\GG\neq0$ is given by
\begin{align}
\delta v_{\nk}|_{\GG\neq0}^\text{filled} \approx \sum_{\QQ\neq0}\frac{\pi}{2c^3|\GG|^3}R_{\nk}^\text{filled}.
\end{align}

\bibliographystyle{apsrev4-1}
\bibliography{refs}